\documentclass[aps,pre,twocolumn,superscriptaddress,showpacs,floatfix]{revtex4-1}

\usepackage{graphicx}
\usepackage{amssymb,amsmath,amsbsy,amsfonts,bm}
\usepackage{epsfig}
\usepackage{multirow}
\usepackage{tikz}
\usepackage{hyperref}

\usepackage{color}

\begin{document}
\title{Layer topology of smectic grain boundaries}

\author{Ren\'e Wittmann}\email{rene.wittmann@hhu.de}
\affiliation{Institut f\"ur Theoretische Physik II: Weiche Materie, Heinrich-Heine-Universit\"at D\"usseldorf, 40225 D\"usseldorf, Germany}
\affiliation{Institut für Sicherheit und Qualität bei Fleisch, Max Rubner-Institut, 95326 Kulmbach, Germany}

\begin{abstract}
Grain boundaries in extremely confined colloidal smectics possess a topological fine structure with coexisting nematic and tetratic symmetry of the director field. An alternative way to approach the problem of smectic topology is via the layer structure, which is typically more accessible in experiments on molecular liquid crystals. Here, we consider exemplary density functional theory results for two-dimensional hard rods to illustrate how the concept of endpoint defects, which appear as tetratic disclinations of quarter charge in director topology, translates to smectic layer topology built around the networks formed by density maxima and minima. By doing so, we elaborate on further advantages of a topological concept evolving around the layer structure rather than the director field, such as providing insight in the structure of edge dislocations or virtual defects at the confining walls.
\end{abstract}

\maketitle

\section{Introduction}
A number of lectures at the 16th European Conference on Liquid Crystals were devoted to
topological defects of smectic liquid crystals \cite{zappone2022one,wittmann2023colloidal,o2023templated},
which can be utilized, for example, to guide the assembly of nanoparticles \cite{jeridi2022unique}.
For a two-dimensional system of colloidal smectics in extreme confinement, the predominant appearance of defects is in the form of grain boundaries, along which the orientational order is frustrated \cite{monderkamp2021topology,annulus}.
Hence, these spatially extended defects possess a topological charge, which can be identified in the same way as that of nematic point disclinations.
Upon superimposing a tetratic symmetry, a topological fine structure of the smectic grain boundaries in terms of quarter-charged point defects becomes visible  \cite{monderkamp2021topology}.
Moreover, well-behaved three-dimensional systems can be studied by their two-dimensional defect structure of representative cross-sections \cite{monderkamp2022topological}.
This classification scheme can be both employed within continuum theories \cite{xia-2021-article,wittmann2023colloidal,paget-2022-article,paget2023complex} and directly applied to structures observed in colloidal experiments \cite{CollLQinSqConf,monderkamp2021topology,wittmann2023colloidal,jull2023curvature}.

Quite generally, a topological fine structure of a spatially extended disclination can become manifest when two different symmetries are superimposed.
As a pedagogical example, let us assume we describe a nematic liquid crystal in terms of a polar vector field, instead of the higher-symmetric nematic tensor field reflecting the actual particle symmetry.
While such a simplified approach is useful in many cases, its limitations become apparent for pairs of half-charged disclinations, which then would be effectively connected by a grain boundary.
Using tensors (or, generally speaking, basing our analysis on a higher symmetry) is thus more instructive as it allows to focus on point defects.
In the present application to smectic grain boundaries, the local tetratic symmetry assumed for the frustrated system does not reflect the particle symmetry (which is nematic),
but the benefit of recovering point disclinations with respect to the higher (tetratic) symmetry remains.

Despite the experimental observation of grain boundaries in cross-sections of thin smectic films with different anchoring conditions \cite{michel2004optical,michel2006structure,coursault2016self,zappone2020analogy,zappone2022one},
the topological fine structure of disclinations in molecular liquid crystals has not yet been investigated.
One reason for this gap could be that the molecular orientation and thus the director field underlying the classification scheme from Ref.~\cite{monderkamp2021topology} is not easily accessible in such experiments.
Instead, the common observable in experiments on smectic phases of molecular liquid crystals is their layer structure.
Accordingly, the historic evolution of smectic topology started by considering the layers as the elementary unit of the system \cite{poenaru1981some,exp_studies,chen2009symmetry,kamien2016topology,aharoni2017composite,machon2019}.
Hitherto, it is not clear how the recent insights on tetratic quarter charges \cite{monderkamp2021topology} fit into a topological picture appealing to smectic layering.

Here, we address the topological charge and fine structure of smectic grain boundaries in (quasi-) two-dimensional colloidal systems from the perspective of the smectic layers.
In doing so, we not only show that all features of director topology can be reproduced by analyzing networks of layers (density maxima) and half-layers (density minima), but we also highlight structures whose classification is more straightforward when focusing on defects in the layering.
We thus provide a topological framework that could be useful for analyzing grain boundaries in thermotropic smectic liquid crystals.

\section{Smectic topology}

The topological analysis of frustrated systems requires in general the identification of defects, at which a certain type of order changes discontinuously.
These defects can be assigned a topological charge, which allows to associate conservation and addition laws analogous to electrodynamics.
In particular, defects with opposite charges can, in principle, annihilate, leaving a structure that is (locally) free of defects.

Topological charges of disclinations can be defined within director topology as $Q=\Delta\phi/2\pi$, where $\Delta\phi$ is the angle of net rotation of the director field upon once traversing a closed contour in counter-clockwise direction along a defect-free region surrounding the defect core~\cite{nem_defects}.
This concept is well established for nematics, for which $\Delta\phi$ must be a multiple of $\pi$,  and thus directly generalizes to smectics, which possess the same orientational symmetry.
In the case of tetratic order, $\Delta\phi$ can generally assume multiples of $\pi/2$.
Thus, upon superimposing a tetratic symmetry, it is possible to choose contours cutting through a grain boundary, which allows to define quarter-charged tetratic point defects with $Q_\text{e}=\Delta\phi/2\pi$ at the ends of smectic grain boundaries \cite{monderkamp2021topology}.
It is important to note that associating such a tetratic fine structure to a grain boundary is not an actual decomposition of the nematic charge.
The distinction between nematic and tetratic disclinations must be kept in mind at all times, as the notion of what is a defect differs for each underlying symmetry, leading to independent charge conservation laws.

Smectic layer topology concerns both defects in positional order (dislocations) and orientational order (disclinations).
The former are typically characterized in terms of their Burgers vector (which is a scalar for smectics \cite{aharoni2017composite}), measuring the amount of layers terminating or new layers formed, i.e., the strength of the dislocation.
Disclinations can, in principle, be defined analogously to director topology, as the layers are largely perpendicular to orientational director.
Here, we evoke a network topological approach to characterize the smectic layering, which provides a comprehensive picture of dislocations and disclinations \cite{hocking2022topological} and is based upon two central pillars.
First, the networks representing the layer structure can be decorated with topological charges $q=1-d/2$ associated with each vertex of degree $d$.
Second, it is not sufficient to only consider the network of the smectic layers (location of the molecular centers), as it is crucial to also account for the so-called half-layers (regions void of molecules) in between \cite{chen2009symmetry,aharoni2017composite,machon2019}.

 In the following, we illustrate different defect structures via the density profiles of hard discorectangles (two-dimensional spherocylinders)  in circular or annular confinement, obtained in Ref.~\cite{annulus} using classical density functional theory~\cite{Evans1979} for two-dimensional hard rods~\cite{wittmann2017phase}.
In particular, we show how the concepts of a topological fine structure and quarter charges translate from director topology to network topology of smectic layers.

\begin{figure}[t]
\begin{center}
\vspace{0.5cm}
\includegraphics[width=0.985\linewidth]{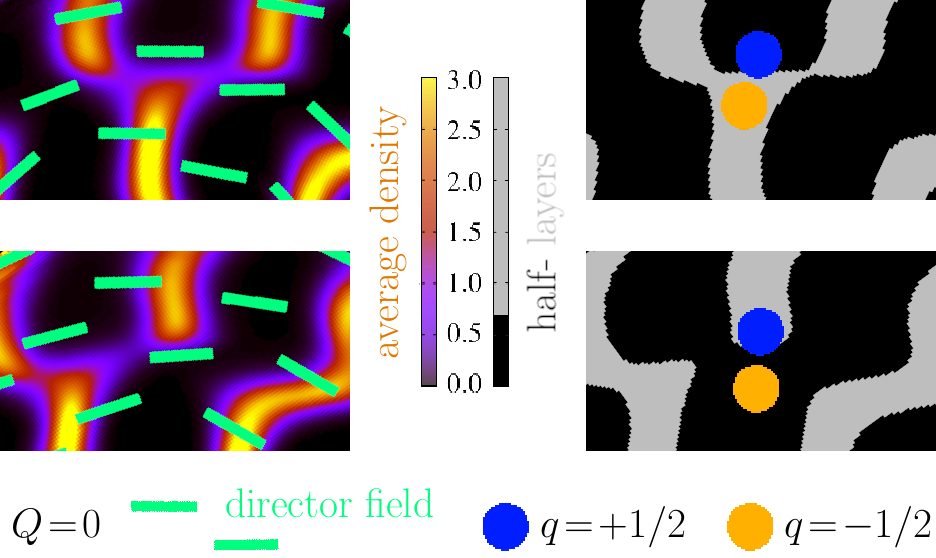}
\caption{\textbf{Edge dislocation from director and layer topology.}
We show an edge dislocation in the form of a branching layer (top) and in the form of a branching half-layer (bottom).
Data are obtained from Ref.~\cite{annulus} and correspond to hard discorectangles with unit length $L$ and aspect ratio $p=10$, forming a Shubnikov state in annular confinement, realized with a hard wall of outer radius $R_\text{out}=3.1L$ and inner radius $R_\text{in}=0.3R_\text{out}$.
\textbf{Director topology:} the plots on the left-hand side depict excerpts of density profiles, where the average center-of-mass density (normalized by the particle volume) is indicated by the color bar and the orientational director field is indicated by green stripes.
As the latter has no discontinuities in the vicinity of an edge dislocation, it possesses no net topological charge $Q=0$.
\textbf{Layer topology:} the plots on the right-hand side depict the same excerpts but without director field and a binary color code highlighting the layers (high density, gray) and half-layers (low density, black).
In addition, we highlight positive (blue) and negative (yellow) topological defects
in the layer- and half-layer networks.
An edge dislocation can thus be recognized by a pair of oppositely charged defects sitting on distinct networks.
\label{fig_ed} }
\end{center}
\end{figure}

\begin{figure*}[t]
\begin{center}
\vspace{0.5cm}
\includegraphics[width=0.885\linewidth]{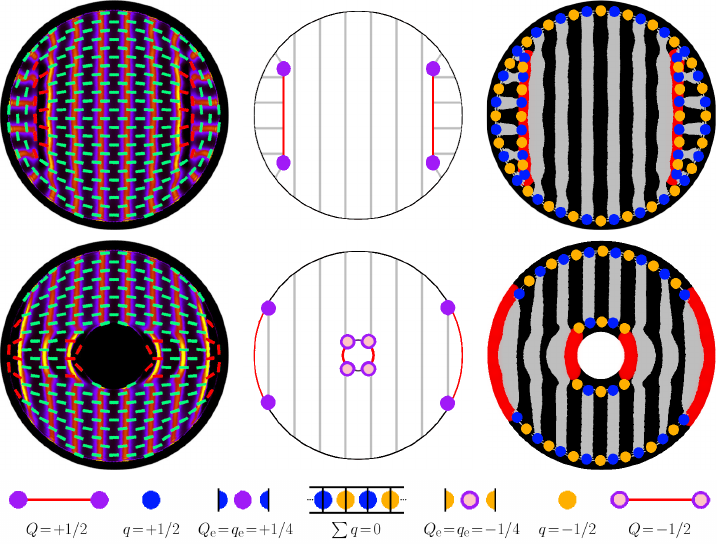}
\caption{\textbf{Grain boundaries from director and layer topology.}
We show a bridge state in annular confinement with two interior grain boundaries (top) and a laminar state in circular confinement with four virtual grain boundaries at the system boundaries (bottom).
Data are obtained from Ref.~\cite{annulus} and correspond to hard discorectangles with unit length $L$ and aspect ratio $p=10$. The radius of the outer hard wall is $R_\text{out}=5.2L$ in each case and the radius of the inner hard wall of the annulus is $R_\text{in}=0.3R_\text{out}$.
\textbf{Director topology:} the plots on the left-hand side depict density profiles as in Fig.~\ref{fig_ed}, where we additionally highlight the discontinuous director field around the grain boundary by a red color.
The illustrations in the center schematically show the layers (gray) and grain boundaries (red), whose topological charge $Q$ is noted in the legend, according to the tetratic quarter charges $Q_\text{e}$ at the endpoints.
\textbf{Layer topology:} the plots on the right-hand side depict the same density profiles with a binary color code as in Fig.~\ref{fig_ed}, where we again show all topological defects $q$ on the binary network, noting that the system boundary (at which no particles are located) is formally part of the half-layers.
In addition, we highlight the grain boundaries by a red background.
As indicated in the legend, oppositely charged neighboring defects average out along system boundaries and grain boundaries,
which allows to identify quarter-charged endpoint defects $q_\text{e}$, being fully consistent with $Q_\text{e}$ from director topology.
\label{fig_GB} }
\end{center}
\end{figure*}

\subsection{Edge dislocations}
To provide a typical example for the identification of network charges and their interpretation, let us first discuss the stability of edge dislocations from a topological point of view, see Fig.~\ref{fig_ed}.
An edge dislocation is a spatial distortion of the layer structure, which does not involve a distortion of the orientational order (contrasting a dislocation from a disclination).
In other words, there is no topological charge associated with an edge dislocation from the point of view of director topology (left panels of Fig.~\ref{fig_ed}).

However, we can identify topological charges in the dual layer network (right panels of Fig.~\ref{fig_ed}):
 layers branch out, which is associated with a negative $q=-1/2$ defect and implies that one half-layer must terminate at this junction, forming the associated positive $q=+1/2$ defect (top row in Fig.~\ref{fig_ed}).
 Vice versa, if a layer terminates on a $q=+1/2$ defect, the opposite $q=-1/2$ defect sits on the branching half-layer (bottom row in Fig.~\ref{fig_ed}).
These observations lead to  the crucial physical interpretation that edge dislocations are topologically stable composite units because oppositely charged defects sit on different networks and may thus not annihilate \cite{aharoni2017composite,machon2019}, despite the zero net charge.
In contrast, the director field is uniform around the dislocation (left panels of Fig.~\ref{fig_ed}) and no charge can be measured at all,
which is certainly consistent, but does not provide any additional insight.

We see from this brief comparison that layer topology brings certain advantages compared to director topology.
Before turning to the topological analysis of  grain boundaries (disclinations) as our main objective,
we comment on a few further aspects of network topology for edge dislocations.
First, for a more profound analysis, the Burgers vector can be inferred from the number of intact layers between the network disclinations, see Ref.~\cite{hocking2022topological} for a detailed discussion.
Second, it is possible to determine the number of edge dislocations in a confined system by counting the interior (i.e., not at the system boundary) network charges \cite{jull2023curvature}.
Finally, it is, in principle, irrelevant whether an edge dislocation is depicted as a branching layer or a new layer forming (branching half-layer), as there is no mathematical distinction.
However, the density fields from particle resolved density functional theory, shown in Fig.~\ref{fig_ed},
implicitly account for the (rod-like) particle shape and thus provide distinct physical pictures of the actual colloidal system, involving certain degrees of branching.
This resolves the ambiguity (if any) in locating the network charges of an edge dislocation.
The only arbitrary input is the choice of the threshold value to distinguish between layers and half-layers.

\subsection{Grain boundaries}
Now we turn to smectic grain boundaries forming in extremely confined systems \cite{monderkamp2021topology}.
To demonstrate the basic topological concepts, Fig.~\ref{fig_GB} shows two characteristic structures, as classified in Ref.~\cite{annulus}:
a bridge state with two interior grain boundaries in circular confinement (top row) and a laminar state with four virtual grain boundaries in annular confinement (bottom row).
As no other singular objects are present, the total charge ($\sum Q=\sum q=1$ for the disk and $\sum Q=\sum q=0$ for the annulus) must be distributed over these grain boundaries.

\subsubsection{Director topology}
Regarding the director field, shown as green stripes in the left panels of Fig.~\ref{fig_GB}, the topology of the grain boundaries can be directly determined from following appropriate contours, encircling the whole region with discontinuous orientations (highlighted in the director field by using a red color instead of green).
In addition, the virtual defects at the confining walls are identified by the deviation from the preferred alignment of the hard rods parallel to the confining walls.
Thus, they can be treated analogously to interior defects by a virtual extension of the director field satisfying this presumed anchoring condition \cite{monderkamp2021topology,annulus}
(in fact, using density functional theory, the virtual defects are directly visible in the director field, as there is always a small probability for each allowed orientation).

As illustrated in the central panels of Fig.~\ref{fig_GB},
the two interior grain boundaries in the bridge state, as well as the two virtual ones at the outer walls in the laminar state, carry a charge of $Q=+1/2$ each,
while those close to the inclusion in the laminar state have $Q=-1/2$.
Since by definition, the nematic order is singular throughout the grain boundary, no further decomposition of this disclination is possible.
Analyzing, however, the director field as being tetratic, such that a  perpendicular orientation of two rods does not represent a defect,
 a fine structure with the isolated charges $Q_\text{e}=+1/4$ or $Q_\text{e}=-1/4$ at the endpoints of positively or negatively charged grain boundaries, respectively, can be revealed and visualized \cite{monderkamp2021topology}.
We demonstrate below that this established picture is consistent with the conclusions that can be drawn from layer topology and present additional insight from the synergy of these two different points of view.

\subsubsection{Layer topology}
In the right panels of Fig.~\ref{fig_GB}, we identify the network topological charges of the grain boundaries in our two confined systems.
 As charge conservation requires a boundary rule (in analogy to the presumed parallel director alignment at the confining walls),
the boundary itself must be associated with one of the two networks~\cite{monderkamp2023network}, i.e., the network of layers, or as done here, the network of half-layers (which intuitively indicate the absence of particles both between the layers and beyond the confining walls).
Doing so, we obtain a sequence of equidistant alternating charges, which average out throughout a defect-free confining wall, as illustrated in the legend of Fig.~\ref{fig_GB}.
As for an edge dislocation, the fact that $\sum q=0$ does not necessarily imply that oppositely charged defects can annihilate, since they are associated with different networks.

To analyze the distribution of layer defects over a grain boundary, let us first take a closer look at the bridge state in the top row of Fig.~\ref{fig_GB}.
The alternating charges at the confining wall indicate that the total charge must be distributed in the interior of the system, i.e., at the two grain boundaries, as also required by director topology (left panel).
Here, these grain boundaries are recognized as the locus of the interior defects in the two networks of layers and half-layers (right panel).
Further inspection reveals again a sequence of alternating charges,  which  ends here  at each side of each grain boundary on a positive defect $q=+1/2$ associated with a terminating layer.
Upon averaging over the defect pairs along the grain boundary, as illustrated in the legend of Fig.~\ref{fig_GB}, we recognize that each endpoint defect has one oppositely charged neighbor which only compensates half of its charge.
The remaining half can be interpreted as an effective point defect with $q_\text{e}=+1/4$ located at the end of the grain boundary.
This leaves the overall picture of a positively charged grain boundary with two quarter charges at the endpoints (central panel),
which is consistent with the charge distribution identified using director topology.

While director topology arguably provides a physically more intuitive picture of the topological fine structure of a grain boundary in terms of coexisting symmetries,
the analogous definition in terms of layer networks directly extends to virtual grain boundaries,
i.e., it is straightforward to identify and locate quarter-charged boundary defects, which are not visible in local order parameter fields \cite{monderkamp2021topology}.
To illustrate this procedure, we consider the laminar structure in the bottom row of Fig.~\ref{fig_GB}, which only possesses one large domain of nearly parallel layers and a practically uniform director field throughout the interior region (left panel).
In this circumstance, a virtual grain boundary can be identified, using layer topology, by the absence of a regular alternating  pattern of oppositely charged defects at the confining walls (right panel).
With this gap, the charges do not average out along the confining walls,
 such that there is a
 positive defect $q=+1/2$, associated with the terminating outermost layers,
 and a negative defect $q=-1/2$, associated with the branching half-layers,
 at each end of each virtual grain boundary at the outer and inner walls, respectively.
Again, averaging over direct neighbors effectively results in the corresponding positive and negative quarter charges $q_\text{e}=\pm 1/4$ at the endpoints (central panel).
While this is also consistent with director topology, the present layer topological approach enables a direct visualization of these virtual endpoint defects.

In summary, the main difference to director topology, where the charge of a grain boundary of any kind follows from considering a closed contour encircling the whole defect, is that an the appearance of interior/virtual grain boundaries in layer topology is different, as it must be identified from the presence/absence of a sequence of alternating network charges.
The quarter charge of the endpoints, on the other hand, directly follows from the exceptional role of the outermost defects in such a sequence, which have only one neighbor.

\section{Conclusions}
To conclude, both layer topology and director topology can be used to identify the topological charge and the fine structure of interior and virtual smectic grain boundaries.
While these two concepts are not strictly dual to each other, the most comprehensive picture can be obtained by employing them in synergy.
(i) The discontinuous order throughout a grain boundary becomes apparent in both the director field and the layer networks.
(ii) The total topological charge of a grain boundary follows when taking into account the full spatial extent, upon observing (iia) the distribution of alternating charges on the dual layer network or (iib) the elongated structure of the nematic defect.
(iii) A fine structure involving quarter charges can be  (iiia) found in layer topology by heuristically splitting and eliminating oppositely charged neighboring network defects and (iiib) substantiated in director topology by identifying point defects in the tetratic field.
(iv) The fine structure of virtual defects, which are located at the confining walls or outside the system, (iva) follows from layer topology in full analogy to interior defects, such that
 this direct observation justifies (ivb) the indirect classification of such defects using director topology, as it is not possible to measure local order parameter fields outside the system.

 In view of practical applications, the layer networks and their defects can be automatically identified and analyzed in both colloidal experiments and simulations of hard rods in two spatial dimensions \cite{monderkamp2023network,jull2023curvature},
  while applications targeting certain three-dimensional systems are within reach \cite{monderkamp2022topological}.
Moreover, the present approach using layer topology can be applied to grain boundaries at which the terminating layers are not perpendicular and to systems in which different physical anchoring conditions are imposed.
This suggests advantages over director topology when targeting applications for molecular liquid crystals \cite{michel2004optical,michel2006structure,coursault2016self,zappone2020analogy,zappone2022one} and related continuum models \cite{ball2023free}.

\section*{Acknowledgements}

The idea for this article originated from conversations with Emmanuelle Lacaze, following stimulating discussions with her and Daniel Beller at the 16th European Conference on Liquid Crystals.
I am  thankful to Randall Kamien for providing valuable insight into layer topology at the Banff International Research Station (Workshop 22w5159).
I also thank Paul A.\ Monderkamp and Hartmut L\"owen for a critical reading of the manuscript and helpful suggestions.

\section*{Funding}

Financial support by the Deutsche Forschungsgemeinschaft (DFG) through the SPP 2265, under Grant No. WI 5527/1-1, is gratefully acknowledged.


\end{document}